\documentclass[12pt]{article}
\usepackage[dvips]{epsfig}
\newcommand{\ee}{$e^+e^-\: $}
\newcommand{\mm}{$\mu^+\mu^-\: $}
\newcommand{\bb}{$b\,\bar{b}\:\; $}
\newcommand{\ssbar}{$s\,\bar{s}\:\; $}
\newcommand{\cc}{$c\,\bar{c}\:\;$}
\newcommand{\qq}{$q\,\bar{q}\:\;$}
\newcommand{\lele}{$l\,\bar{l}\:\;$}
\newcommand{\nn}{$\nu\bar{\nu}\: $}
\newcommand{\mh}{M$_{H} \: $}

\newcommand{\toto}{$t\,\bar{t}\:\; $}

\newcommand{\pfr}{$\rightarrow\:$}
\newcommand{\gaga}{$\gamma \gamma\:\;$}
\newcommand{\noi}{\noindent}
\newcommand{\BS}{\bigskip}

\begin{document}

\pagestyle{empty}

\noi DESY 99 - 109

\BS
\noi July 1999
%\section*{

\vspace{3cm}
\begin{center}
\LARGE{\bf Branching Fraction Measurements of the SM Higgs
 with a Mass of 160 GeV at Future \\
 Linear \ee Colliders}
\end{center}

\vspace{2.0cm}
\large
\begin{center}
E. Boos$^1$, V. Ilyin$^1$, A. Pukhov$^1$, M. Sachwitz$^2$ and
 H.J. Schreiber$^2$  \\
\end{center}

\bigskip \bigskip  
\begin{center}
$^1$ Institute of Nuclear Physics, Moscow State University, \\
119 899 Moscow, Russia \\ [1cm]

$^2$ DESY Zeuthen, 15735 Zeuthen, FRG \\
\end{center}

\newpage
\pagestyle{plain}
\pagenumbering{arabic}

%----------------------------------------------------------------------
% ABSTRACT
%----------------------------------------------------------------------
\section*{Abstract}

Assuming an integrated luminosity of 500 fb$^{-1}$ and a
center-of-mass energy of 350 GeV,
we examine the prospects for measuring branching fractions of a
Standard Model-like Higgs boson with
a mass of 160 GeV at the future linear \ee
collider TESLA when the Higgs is produced
via the Higgsstrahlung mechanism,
\ee \pfr HZ. We study in detail the precisions achievable
for the branching fractions of the Higgs into WW$^*$, ZZ$^*$
and \bb. However,
the measurement of BF(H \pfr \gaga)
 remains a great challence.
Combined with the expected error for the inclusive
Higgsstrahlung production rate the uncertainty
for the total width of the Higgs is estimated.

%----------------------------------------------------------------------
% INTRODUCTION
%----------------------------------------------------------------------
\section{Introduction}

Discovery and study of Higgs boson(s) will be of primary importance at
a next linear \ee collider. After discovery of a Higgs (H), the task will
be to determine as precisely as possible and in a model-independent
manner its fundamental couplings and total width.
 In the Standard Model (SM) \cite{glas},
the mass range from $\sim$100 GeV to about 220 GeV
is favored by comparing precision electroweak data with
SM predictions \cite{erler}. For Higgs masses at 120 and 140 GeV
detailed branching fraction measurement investigations can be found in
e.g. refs.\cite{hil}, \cite{sach}, \cite{mbatetal}. In this paper we assume
a Higgs mass of 160 GeV. For such masses the Higgs decays
mostly into WW$^*$ whereas the \bb decay rate is stronly
suppressed to few percent, opposite to the situation at smaller
masses. Additionally, at \mh $\sim$ 160 GeV
the background expected
to contribute is more complicated since a mixture of WW (where both
W's are on-mass-shell) and WW$^*$ (with one of the W's being off-mass-shell)
exists under the Higgs, and due to detector
resolution effects the observed width of the Higgs boson is expected
to be much larger than its natural total width.
Therefore finite W width effects
have to be accounted for to enable
accurate results.

Future linear \ee colliders operating in the 300 to 500
GeV center-of-mass energy region are ideal machines to investigate the
Higgs sector in this mass regime: besides easily Higgs discovery, all
major decay modes can be explored.

In order to access the actual capability of an \ee collider to an
analysis of the SM Higgs boson it is extremely
appealing to apply recently developed tools, thanks to the effort of
several groups. In particular, we include in this study

\begin{itemize}

\item the full matrix elements for 4-particle final states beyond the
  usual approximation of computing Higgsstrahlung production cross
  sections times branching fractions, using the program package
  CompHEP \cite{boos}. In this way, all irreducible background and possible
  interferences between signal and background diagrams are accounted
  for. Total and partial widths of the Higgs boson corrected for
  QCD and QED loop contributions were implemented in CompHEP and
  cross-checked with the HDECAY program \cite{hdecay}.
  For parton shower and hadronization procedures as well as particle
  decays the PYTHIA/JETSET package \cite{pyt} has been interfaced;

\item initial state QED \cite{kur} and beamstrahlung for the TESLA linear
  collider option \cite{schul};

\item a detector response \cite{pohl}, with detector parameters as designed in
  a series of workshops for the linear \ee collider Conceptual Design
  Report \cite{brink1};

\item all important reducible background reactions expected to
  contribute.
\end{itemize}
The cm energy $\sqrt{s}$ chosen for our study is 350 GeV avoiding thus
\toto pair production as background. Although $\sqrt{s}$ is not
optimized for the best Higgs production rate, it is a compromise
between largest machine luminosity and sufficient event production
for the process
\begin{equation}
e^+e^- \rightarrow  H(160) Z.
\end{equation}
All the results presented are based on the high-luminosity option of
the TESLA machine \cite{brink2} with an instantaneous luminosity
L = 3$\cdot 10^{34}$cm$^{-2}$sec$^{-1}$
resulting to an accumulated luminosity of 500 fb$^{-1}$ within about
two years of running.

%----------------------------------------------------------------------
% sekt. 2.
%----------------------------------------------------------------------
\section{SM Signals, Backgrounds and Branching Fractions}

Within the SM, the Higgs boson with \mh = 160 GeV decays
mostly to WW$^*$, in few percent
of the time to ZZ$^*$ and \bb and
with a very small rate of about 0.5 $\cdot 10^{-3}$ to two photons.
We consider only these decay modes as all others are expected
to be more difficult to measure.
The aim of this study is to provide
the statistical uncertainties of the branching fractions
for the decay modes mentioned by
measuring $\sigma$(\ee \pfr HZ) $\cdot$ BF(H \pfr X)
$\cdot$ BF(Z \pfr Y), $\; \sqrt{S+B}/S$, where S(B) is the number
of Higgs (background) events observed in a small interval of the
invariant mass X, centered around \mh. Branching fraction errors
of the Higgs boson are then computed after convolution with the precision
of the inclusive Higgs production cross section $\sigma$(HZ) of \linebreak
2.8 \% \cite{sach}, \cite{pauolo}. Any Z boson branching fractions
errors are neglected.

%----------------------------------------------------------------------
% sekt. 2.1
%----------------------------------------------------------------------
\subsection{The Branching Fraction BF (H \pfr WW$^*$)}

For \mh = 160 GeV, the Higgs decays mostly to
WW$^*$, i.e. to a final state where one of the W's is on-mass-shell
and the other is off-mass-shell. Since we are interested to make a
signal-to-background analysis as meaningful as possible we also
consistently evaluate the background rates. Most of this background
comes from \ee \pfr WW$^*$Z and WWZ events and requires
an accurate treatment of finite W width effects.

Therefore exact matrix elements are used to calculate
by means of CompHEP the processes
\begin{equation}
e^+e^- \rightarrow \bar{u} d W Z
\end{equation}
\begin{equation}
e^+e^- \rightarrow u \bar{d} W Z
\end{equation}
and
\begin{equation}
e^+e^- \rightarrow WWZ,
\end{equation}
with proper matching below and above the W pair threshold,
and to generate unweigthed events. The final states $\bar{c}$sWZ
and c$\bar{s}$WZ were taken into account by doubling the number
of events for reactions (2) and (3), respectively.
In order to avoid large
not useful event samples
the following cuts were applied to each
final state particle during the generation stage:

\begin{itemize}
\item  energy $>$ 10 GeV ;
\item  polar angle $\Theta > 5^{\circ}$ ;
\item  the invariant mass of any particle pairing M$_{ik} >$ 10 GeV.
\end{itemize}

Due to the large H \pfr WW$^*$ branching fraction and
the luminosity assumed, thousands of signal
events are expected. In the following we restrict our analysis
to only Z \pfr \ee / $\mu^+ \mu^-$  and
W \pfr \qq decays for reasons of simplicity. Therefore,
the signature we have to search for
consists of four jets accompanied by two opposite-charged leptons.

For each
of the reactions (2) - (4) we apply a series of cuts to remove
much of the reducible background and, if
needed, further dedicated criteria are applied to suppress any remaining
irreducible background.

In particular, the cuts we have adapted to select \ee \pfr HZ \pfr
WW$^*$Z events are

\begin{itemize}
\item the visible energy of the event, E$_{vis}$, exceeds 200 GeV;
\item the total transverse energy of the event is larger than 40 GeV;
\item the total momentum along the beam direction
      is restricted to be within $\pm$ 120 GeV;
\item the number of tracks per event is larger than 20;
\item out of all leptons found, at least one \ee or \mm pair has
  an invariant mass within M$_Z \pm$6 GeV;
\item for each jet \footnote {The two leptons selected in the previous
    step were excluded from the jet finder.}
  we require
\begin{itemize}
\item E$_{jet} >$ 10 GeV ;
\item $|$cos $\theta_{jet}| <$ 0.85 ;
\item number of particles /jet $\ge$ 4 ;
\item angle (jet$_i$, jet$_k) >  10^{\circ}$;
\end{itemize}
\item select from all two-jet pairings with 70 GeV $<$ M(\qq) $<$ 90 GeV
   the one best compatible with M$_W$.
\end{itemize}

According to this procedure events
were selected in which the Higgs
recoils against the Z \pfr \lele boson, with H \pfr WW$^*$
and W \pfr \qq decays.

Reducible background events from
\begin{equation}
e^+e^- \rightarrow W^+W^- \rightarrow 4 jets
\end{equation}
\begin{equation}
e^+e^- \rightarrow ZZ \rightarrow 4 jets
\end{equation}
which might mimic our topology and fulfil
the selection criteria
are small. As an example, 150.000 events
of reaction (6) expected for 500 fb$^{-1}$ were
processed and no entry appeared in the final WW$^{(*)}$ mass
distribution after all cuts.

\begin{figure*}[h!t]
\begin{center}
%\mbox{\epsfxsize=17cm\epsfysize=15.5cm\epsffile{fig_3023ww.eps}}
\mbox{\epsfxsize=17cm\epsffile{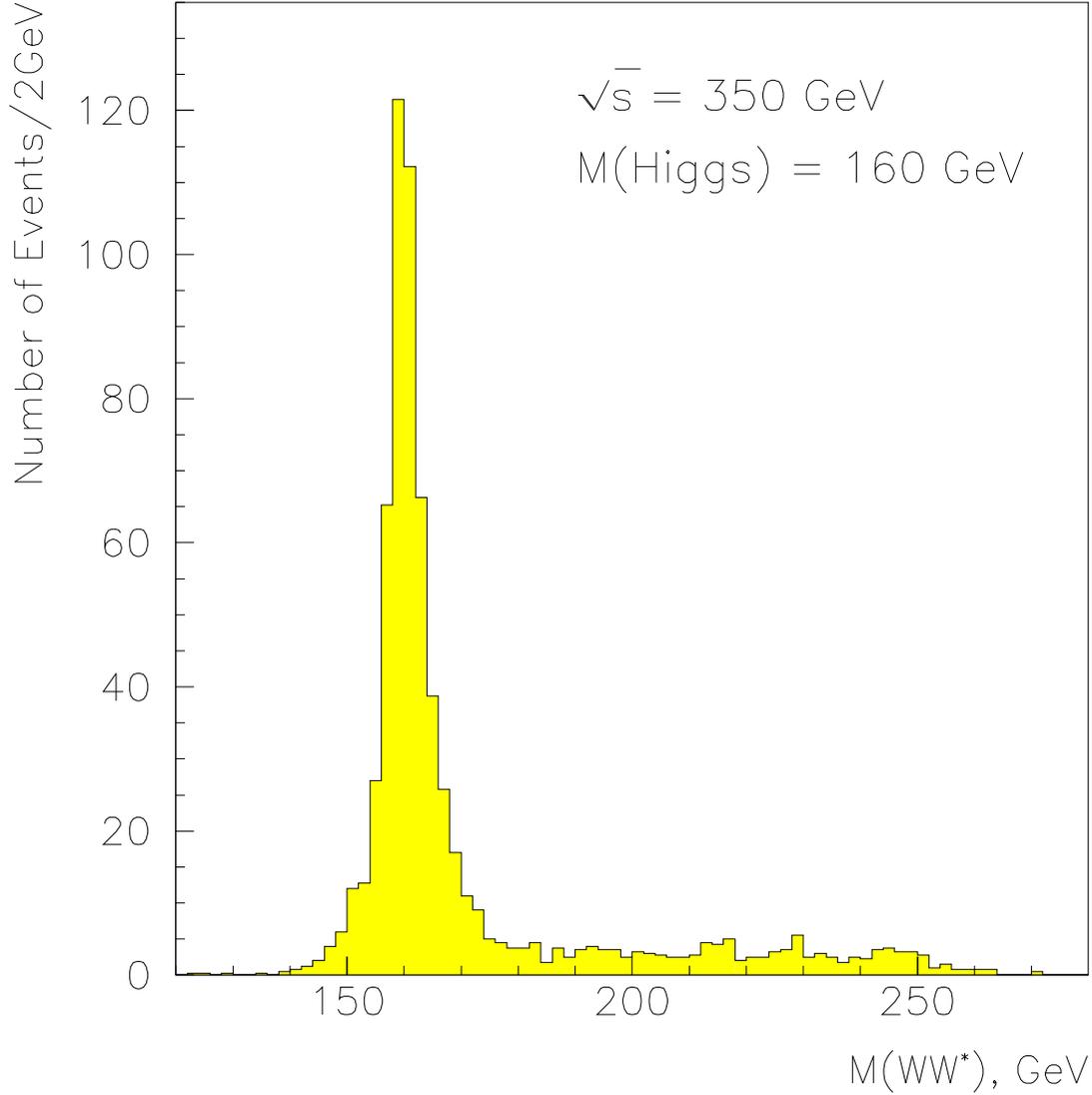}}
\end{center}
\vspace{-0.5cm}
\caption[ ]{\sl The combined WW$^{(*)}$ mass distribution
 after signal and background event selection.}
\vspace{8mm}
\label{fig:fig_3023ww}
\end{figure*}

Fig.\ref{fig:fig_3023ww} shows the WW$^{(*)}$ mass spectrum
for the surviving events.
% M(WW)
%------------------------------------------------------------------
A clear Higgs signal is visible over a very small background which is
mainly due to non-signal diagrams contributing
in reactions (2) and (3).
We estimate a statistical error
of 2.5 \%  for  $\sigma$(HZ) $\cdot$
BF(H \pfr WW$^*$) $\cdot$ BF(Z \pfr \lele). After
convolution with the error of $\sigma$(HZ) the precision
for BF(H \pfr WW$^*$) is found to be $\pm$3.5 \%.
It is worth to mention that
\ee \pfr HZ \pfr (ZZ$^*$) (\lele) \pfr 4 jets + \ee/\mm events,
which have the same signature and could contribute to
the final WW$^{(*)}$ mass spectrum,
were subtracted from the distribution in Fig.1.

%----------------------------------------------------------------------
% sekt. 2.2
%----------------------------------------------------------------------

\subsection{The Branching Fraction BF (H \pfr ZZ$^*$)}

Information for the H \pfr ZZ$^*$ branching fraction can be
obtained either from the
inclusive Higgs production cross section $\sigma$(HZ) which is
proportional to the ZZH coupling-squared or from studies of ZZ$^*$ mass
distributions. The latter spectra, obtained from the reaction
\begin{equation}
e^+e^- \rightarrow ZZ^* Z,
\end{equation}
contribute to four event topologies depending
 on the Z decay modes selected:

\begin{itemize}
\item 6-jet events;
\item 4-jet plus two opposite-charged lepton events;
\item 2-jet plus two pairs of opposite-charged lepton events;
\item three pairs of opposite-charged lepton events.
\end{itemize}

Since large event rates are expected
from the background reaction \ee \pfr WWZ \pfr 6 jets and the signal
process \ee \pfr HZ \pfr WW$^*$Z \pfr 6 jets, the 6-jet topology has
been discarded from our study. Also the
three lepton pair topology had been omitted
due to its very small event rate.

We first consider the 2-jet 4-lepton topology. By means of CompHEP
the following reactions
\begin{equation}
e^+e^- \rightarrow d\bar{d} ZZ
\end{equation}
\begin{equation}
e^+e^- \rightarrow u\bar{u} ZZ
\end{equation}
\begin{equation}
e^+e^- \rightarrow ZZZ
\end{equation}
were generated and properly weighted to account
for \ssbar, \cc and \bb contributions in reactions (8) and (9).
The two Z bosons (and for reaction (10), two Z's out of the three) are
allowed to decay to electron respectively muon pairs. After event
reconstruction we require besides
some more general cuts

\begin{itemize}
\item only \ee/\mm, \ee/\ee or \mm/\mm lepton combinations
  in the final state;
\item $|$cos $\theta_l| <$ 0.95 for each lepton;
\item there exists at least one lepton pair

\begin{itemize}
\item with M$_{ll} $ = M$_Z \pm$6 GeV;
\item E$_l$ + E$_{\bar{l}} >$  125 GeV;
\item which recoils against the (\qq\lele\hspace{-1mm}) system with
 an invariant mass of less than 220 GeV;
\end{itemize}

\item two and only two jets in the final state,
  after excluding the four leptons from the jet finder;
\item for each jet we demand

\begin{itemize}
\item E$_{jet} >$ 10 GeV;
\item $|$cos $\theta_{jet}| >$ 0.9,
\item number of particles/jet $\ge$ 4;
\item angle (jet$_1$, jet$_2$) $>$ 20$^{\circ}$.
\end{itemize}
\end{itemize}

\begin{figure*}[h!t]
\begin{center}
%\mbox{\epsfxsize=17cm\epsfysize=15.5cm\epsffile{fig_7022_zllll.eps}}
\mbox{\epsfxsize=17cm\epsffile{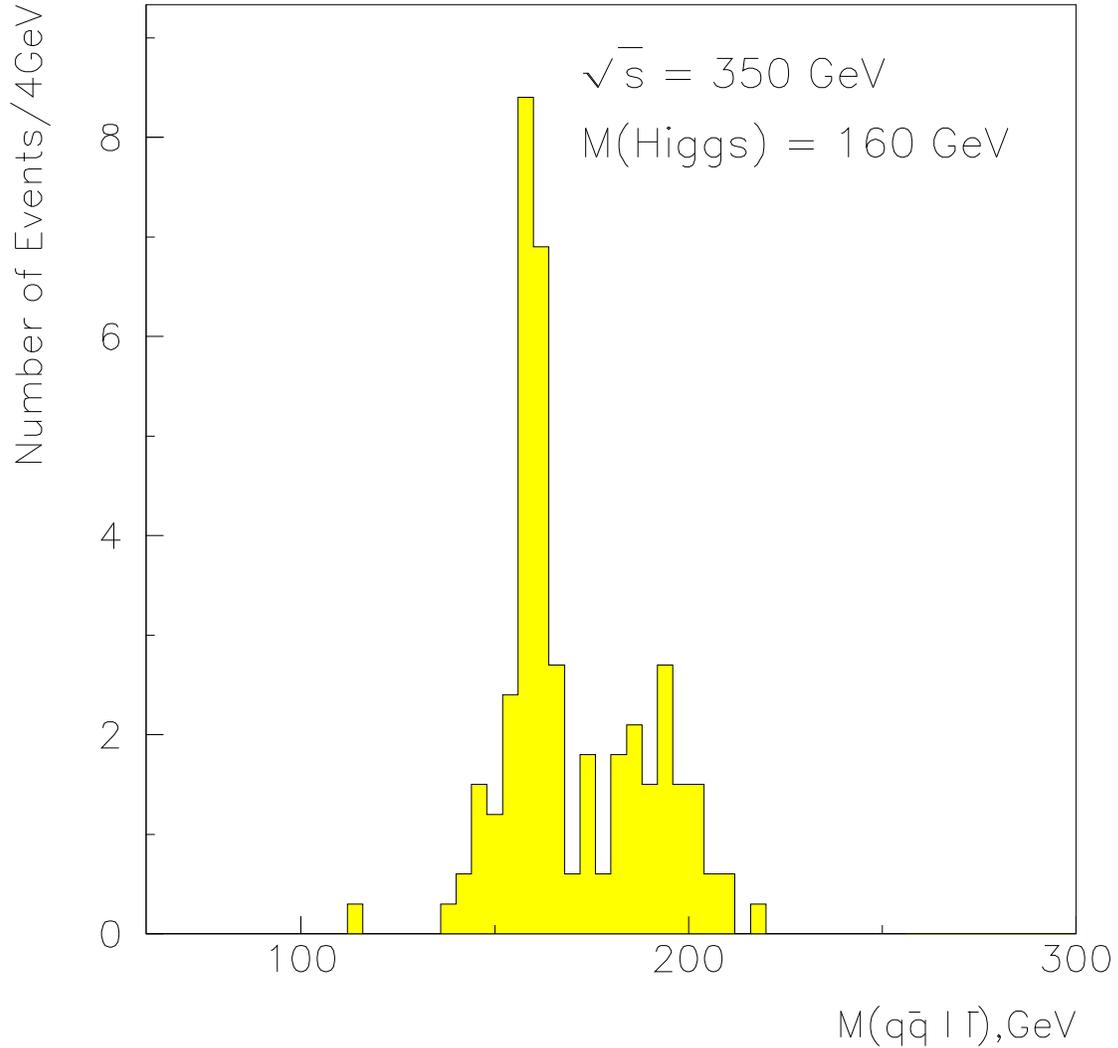}}
\end{center}
\vspace{-0.5cm}
\caption[ ]{\sl The ZZ$^*$ \pfr \qq \lele mass distribution from the
  selected events of the 2-jet 4-lepton topology.}
\vspace{8mm}
\label{fig:fig_7022_zllll}
\end{figure*}

The resulting M(\qq\lele\hspace{-2mm}) mass distribution is
shown in Fig.\ref{fig:fig_7022_zllll}.
% M(ZZ) agaist 2 leptons
%------------------------------------------------------------------
Higgs boson production is clearly observed on very small background.

\vspace{0.5cm}
The 4-jet 2-lepton topology, our second candidate to search for
H \pfr ZZ$^*$ decays, is expected to involve
about 10 times more signal events due to the large Z \pfr \qq
decay rate. This advantage is however to a great extent
compensated by demanding stronger
selection criteria to reduce the larger background expected. The
procedure applied to select signal from background events
requires

\begin{itemize}
\item E$_{vis} >$ 260 GeV;
\item the total transverse energy is restricted between 50 and 290
  GeV;
\item the total momentum along the beam line is within $\pm$ 80 GeV;
\item only \ee or \mm pairs are accepted;
\item the number of jets, excluding the two leptons, is four;
\item for each jet we require

\begin{itemize}
\item E$_{jet} >$ 10 GeV;
\item $|$cos $\theta_{jet}| >$ 0.9;
\item number of particles/jet $\ge$ 6;
\item ange (jet$_i$, jet$_k$) $> 15^{\circ}$;
\end{itemize}

\item the two-jet invariant mass is within M$_Z \pm$10 GeV;
\item the dilepton invariant mass is within M$_Z \pm$6 GeV;
\item E$_l$ + E$_{\bar{l}} >$ 125 GeV;
\item the mass recoiling against the Z (with Z \pfr \qq)
 does not exceed 200 GeV.
\end{itemize}

According to these criteria we try to select \ee \pfr HZ
\pfr (ZZ$^*$)Z \pfr (\lele\qq)(\qq) events,
where the Z boson against the Higgs decays hadronically to \qq.

\begin{figure*}[h!t]
\begin{center}
%\mbox{\epsfxsize=18cm\epsfysize=17.0cm\epsffile{fig_7028_zllz.eps}}
\mbox{\epsfxsize=17cm\epsffile{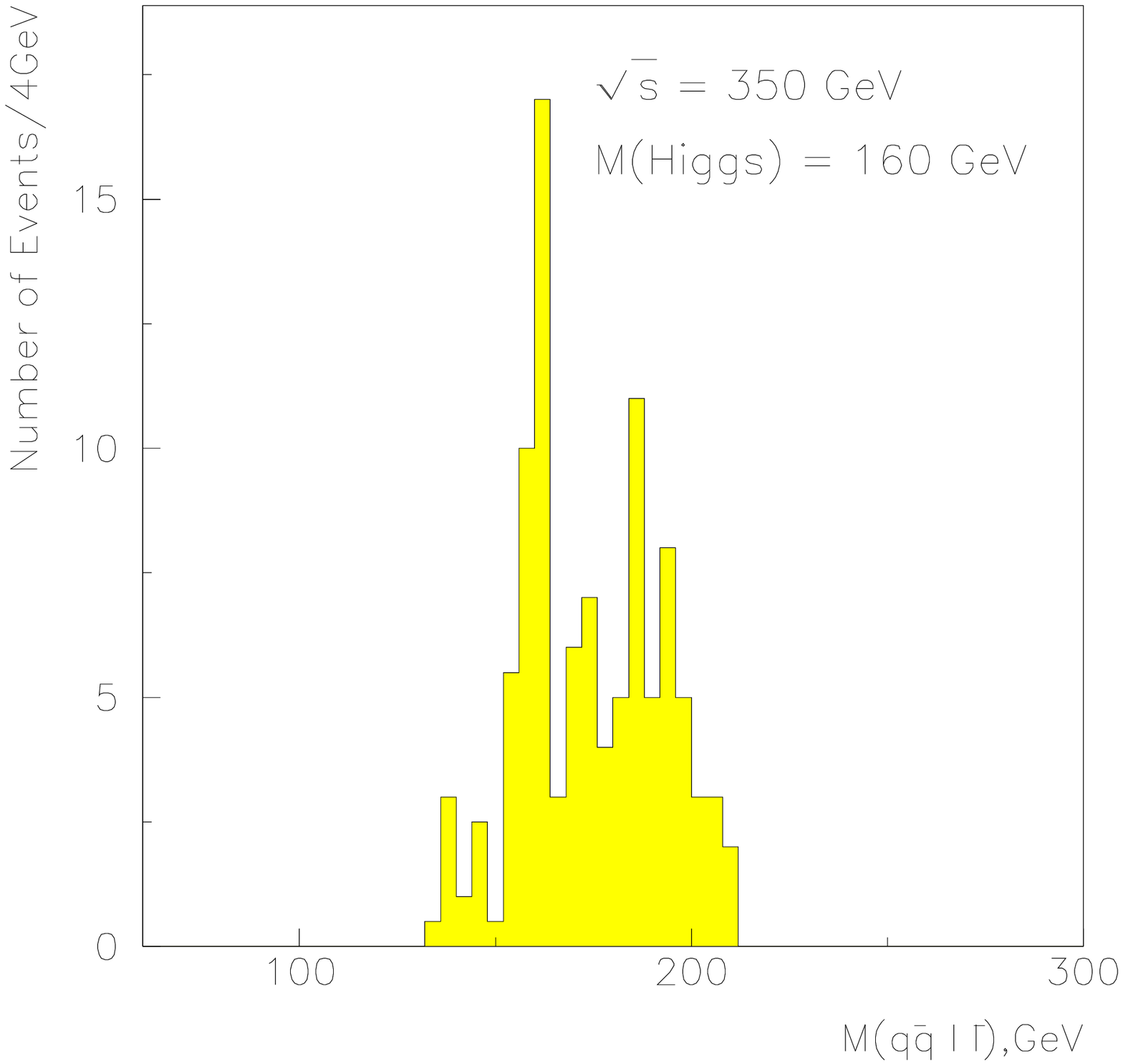}}
\end{center}
\vspace{-0.5cm}
\caption[ ]{\sl The ZZ$^*$ \pfr \qq \lele mass distribution from the
  selected events of the 4-jet 2-lepton topology.}
\vspace{6mm}
\label{fig:fig_7028_zllz}
\end{figure*}

In the final (\qq\lele\hspace{-2mm}) mass spectrum,
shown in Fig.\ref{fig:fig_7028_zllz}, a convincing
Higgs signal can be seen on some remaining background which peaks
near 190 GeV due to surviving \ee \pfr ZZ$^{(*)}$Z events.
% M(ZZ) Z--> ll, H agaist q q-bar
%------------------------------------------------------------------
The combined spectrum of Figs.2 and 3 yields for $\sqrt{S+B}/S =
\Delta [\sigma$(HZ) $\cdot$ BF(H \pfr ZZ$^*$)] = 14.2 \%, which results
to a statistical error of $\pm$14.5 \% for
BF (H \pfr ZZ$^*$), after convolution with
the uncertainty of $\sigma$(HZ).

%----------------------------------------------------------------------
% sekt. 2.3
%----------------------------------------------------------------------

\subsection{The Branching Fraction BF (H \pfr \bb\hspace{-1mm})}

In contrast to the Higgs mass region below $\sim$130 GeV with the
dominant H \pfr \bb decay mode , heavier Higgs bosons are predicted
to have a small or negligible decay rate
to \bb. For M$_H$ = 160 GeV, the SM predicts a
3.8 \% branching fraction which requires
high luminosity for a precise measurement.

The topologies expected for \ee \pfr HZ \pfr (\bb\hspace{-1mm}) Z production
consist either of 4-jet events involving two b-jets
\begin{equation}
e^+e^- \rightarrow b\bar{b}\hspace{2mm} q\bar{q}
\end{equation}
or two b-jet events accompanied by a pair of opposite-charged leptons
\begin{equation}
e^+e^- \rightarrow b\bar{b}\hspace{2mm} e^+e^-/\mu^+\mu^- .
\end{equation}
Events with Z \pfr \nn decays are excluded from the analysis since we
demand visible Z decays.

Most of the background comes from the channels
\begin{equation}
e^+e^- \rightarrow q\bar{q}\hspace{2mm} q\bar{q},
\end{equation}
and 
\begin{equation}
e^+e^- \rightarrow  q\bar{q}\hspace{2mm}  e^+e^- / \mu^+\mu^-,
\end{equation}
with q = u, d, s and c.

Our criteria to select events of reaction (11) are the following

\begin{itemize}
\item four and only four jets occur in the final state;
\item for each jet we require

\begin{itemize}
\item E$_{jet} >$ 12 GeV;
\item  $|$cos $\theta_{jet}| <$ 0.85;
\item angle (jet$_i$, jet$_k$) $> 10^{\circ}$;
\item number of particles/jet $\ge$ 8;
\end{itemize}

\item out of the four jets demanded two jets are tagged as b-jets;
  a jet is defined as a b-jet if
\begin{itemize}
\item the number of charged particles with large impact parameter
 is $\ge$ 3 for one jet and $\ge$ 4 for the other;
\item a track is considered as a 'large impact parameter track' if
  its distance of closest approach to the primary vertex in the
  (r,phi) or the (r,z) projection is $\ge$ 3, in units of its error.
\end{itemize}

\item the two remaining jets have an invariant mass
  within M$_Z \pm$10 GeV.
\end{itemize}

\begin{figure*}[h!t]
\begin{center}
%\mbox{\epsfxsize=17cm\epsfysize=15.5cm\epsffile{fig_5032bbqq.eps}}
\mbox{\epsfxsize=17cm\epsffile{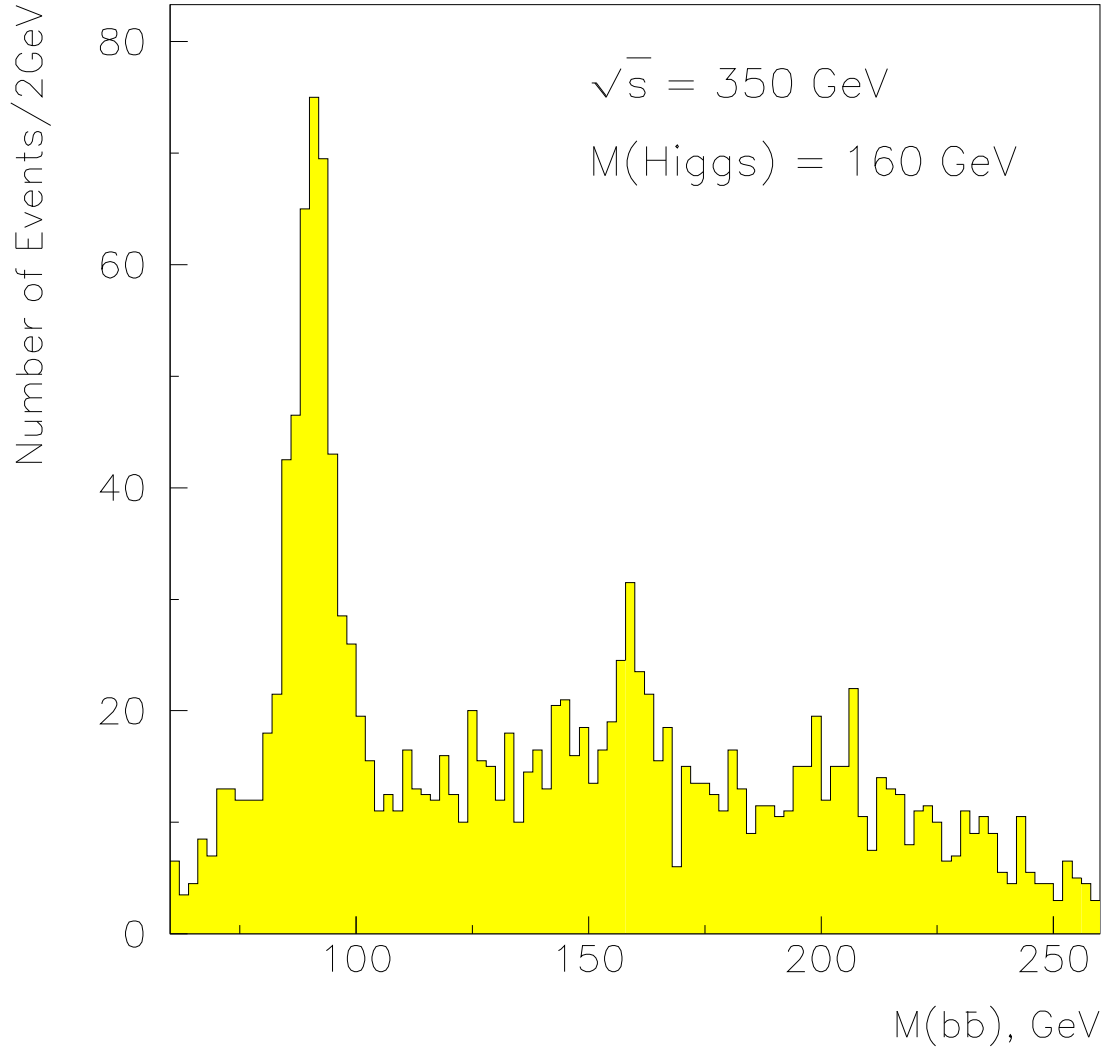}}
\end{center}
\vspace{-0.5cm}
\caption[ ]{\sl The \bb mass distribution from the selected events
 of the 4-jet topology.}
\label{fig:fig_5032bbqq}
\end{figure*}

The resulting b-tagging efficiency turns out to be $\sim$ 65 \% while
the probability for a light \qq pair to be tagged
as a \bb pair is below 3 \%.
For the surviving events the \bb invariant mass spectrum is shown in
Fig.\ref{fig:fig_5032bbqq}. Besides the dominating Z boson, the Higgs
% M(bb) Z--> q q-bar
%------------------------------------------------------------------
in its \bb decay mode is clearly visible at 160 GeV. An estimate of the
signal and background rates in the vicinity of M$_H$
results to an error of $\sigma$(HZ) $\cdot$ BF(H \pfr \bb) = $\pm$16 \%.

\vspace{5mm}
Events of the 2-jet dilepton signal topology, produced about ten times less
frequently, are selected by
\begin{itemize}
\item identifying an \ee or \mm pair with an invariant mass
  M$_Z \pm$6 GeV;
\item two and only two jets, excluding the lepton pair
 from the jet finder,
 are tagged as b-jets, with properties as defined above.
\end{itemize}

\begin{figure*}[h!t]
\begin{center}
%\mbox{\epsfxsize=17cm\epsfysize=15.5cm\epsffile{fig_4032bbll.eps}}
\mbox{\epsfxsize=17cm\epsffile{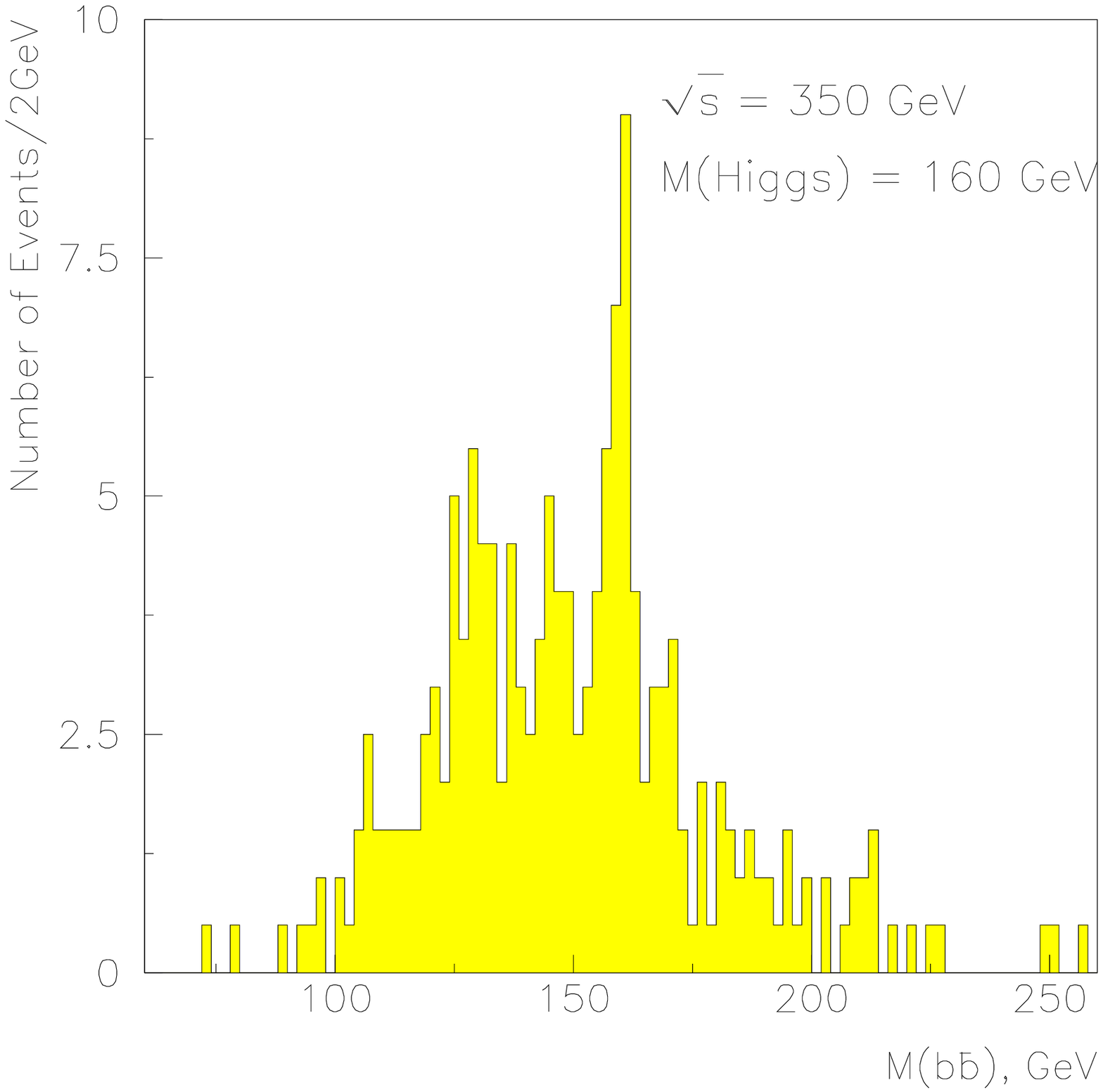}}
\end{center}
\vspace{-0.5cm}
\caption[ ]{\sl The \bb mass distribution from the selected events
 of the 2-jet dilepton topology.}
\label{fig:fig_4032bbll}
\end{figure*}

The resulting \bb mass distribution is
shown in Fig.\ref{fig:fig_4032bbll} and, if combined
with the spectrum of Fig.4, the statistical error for
BF(H \pfr \bb) results to $\pm$12 \%.
% M(bb), Z--> l l
%------------------------------------------------------------------

%----------------------------------------------------------------------
% sekt. 2.4
%----------------------------------------------------------------------
\subsection{The Branching Fraction BF (H \pfr \gaga\hspace{-2mm})}

The Standard Model predicts for BF(H \pfr \gaga\hspace{-2mm})
with M$_H$ = 160 GeV
a rate of 0.54 $\cdot 10^{-3}$, so that only very large
statistics experiments combined with an electromagnetic calorimeter of
excellent resolution might have access to this quantity.
Deviations of BF(H \pfr \gaga\hspace{-2mm}) measurements from
the SM expectations would
be of great importance. In particular, by virtue of the fact that the
coupling H \pfr \gaga arises from charged loops, large deviations
from the SM value due to new particles (e.g. a fourth generation,
supersymmetry) are possible. Measuring BF(H \pfr \gaga\hspace{-1mm})
at the next
linear collider will be, regardless of the size of the deviation from
the SM prediction, important to understand the nature
of the Higgs and provides
hints of new physics that may be beyond the Standard Model.

In principle, similar arguments are also valid for Higgs decays like
H \pfr 2 gluons or H \pfr $\gamma$Z. However, their measurability
is expected to be difficult and requires dedicated studies
in future. Here we consider the easily recognizable
H \pfr \gaga decay mode with
the hope to observe the 2-photon Higgs decay despite the presence
of large background.

\begin{figure*}[h!t]
\begin{center}
%\mbox{\epsfxsize=17cm\epsfysize=17.5cm\epsffile{diagr.ps}}

\vspace*{-12.5cm}
\epsfig{file=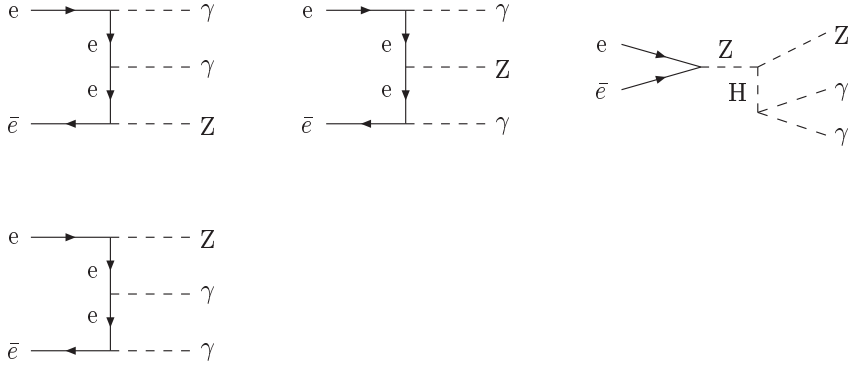,width=18cm}
\end{center}

\vspace{-7.0cm}
\caption[ ]{\sl SM diagrams contributing to reactions (15) and (16).}
\vspace{10mm}
\label{fig:diagr}
\end{figure*}

For an accumulated luminosity of 500 $fb^{-1}$
we only expect 39 events from the Higgsstrahlung process
\begin{equation}
e^+e^- \rightarrow H(160) Z \rightarrow \gamma \gamma\hspace{2mm} q\bar{q},
\end{equation}
which have to be confronted with orders of magnitude larger
irreducible background from the channel
\begin{equation}
e^+e^- \rightarrow \gamma \gamma Z \rightarrow \gamma \gamma\hspace{2mm} q\bar{q}.
\end{equation}
The signal and background diagrams contributing to these reactions
are presented in Fig.\ref{fig:diagr}.
% diagrams
%------------------------------------------------------------------

Additional background comes from the reaction
\begin{equation}
e^+e^- \rightarrow \gamma Z \rightarrow \gamma \hspace{1mm}q\bar{q}\hspace{1mm} (\gamma),
\end{equation}
where the photon in parentheses comes from initial state radiation
and/or quark fragmentation tail and fluctuation effects.
Since initial state photon radiation is already covered 
to a great extent by reaction (16) we have not combined both background
contributions in the following. Hence, reaction (16)
represents only some lower background limit.

In order to remove most of this background while retaining significant
signal events
in the \gaga mass close to M$_H$
dedicated selection procedures are absolutely mandatory.
After exploring a variety of requirements like e.g.

\begin{figure*}[h!t]
\begin{center}
%\mbox{\epsfxsize=17cm\epsfysize=17.5cm\epsffile{fig_1244ggqq.eps}}
\mbox{\epsfxsize=17cm\epsffile{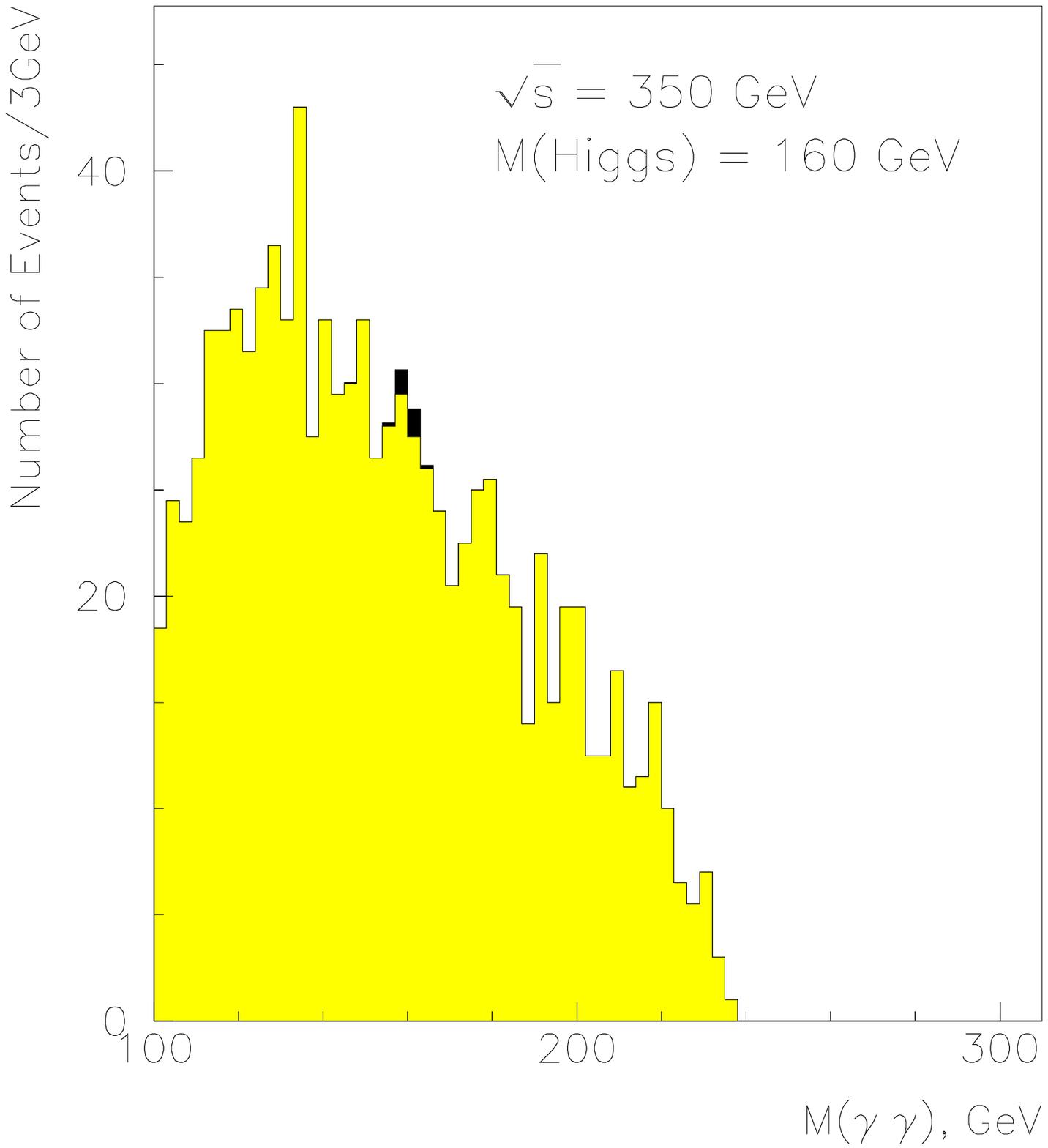}}
\end{center}
\vspace{-0.5cm}
\caption[ ]{\sl The \gaga mass spectrum from the surviving
 signal and background events
 of the reaction \ee \pfr \gaga Z \pfr \gaga \qq. The Higgs events
 are indicated cross-hatched.}
\label{fig:fig_1244ggqq}
\end{figure*}

\begin{itemize}
\item total visible energy $>$ 240 GeV;
\item total transverse energy $>$ 30 GeV;
\item $p_T(\gamma_1) > p_T(\gamma_2) >$ 20 GeV;
\item $p_T(\gamma_1) + p_T(\gamma_2) >$ 64 GeV;
\item $|$cos $\theta_{\gamma_{1/2}}| <$ 0.7;
\item two and only two jets with an invariant mass within M$_Z \pm$10 GeV,
   after removal of the two selected
   photons from the jet finder algorythm;
\item  $|$cos $\theta_{jet}| < $ 0.7 for each jet,
\end{itemize}

we obtain the \gaga invariant mass distribution
as shown in Fig.\ref{fig:fig_1244ggqq}.
% M(ga ga)
%------------------------------------------------------------------
As can be seen, the
background in the Higgs region
is overwhelming. The surviving Higgs event rate,
indicated as cross-hatched in the figure,
is so small that a reasonable BF (H \pfr \gaga\hspace{-3mm}) measurement
remains challenging for future investigations. A multidimensional
analysis based on a likelihood estimator which includes 16 variables
had not improved the results obtained \footnote {In
particular, variables like production and decay angles, transverse
energies, thrust and acoplanarity different for signal and
background events were taken into account, and the resulting
estimator (probability of being a signal event) was required to
be $>$ 0.92.}.
Whether the addition
of H \pfr \gaga fusion events, with an about four times less 
production cross section,
would significantly alter our conclusion remains to be studied
in detail.

%----------------------------------------------------------------------
% sekt. 3
%----------------------------------------------------------------------

\section{The total Higgs width $\Gamma_{tot}$(H)}

Besides the measurements of the Higgs branching fractions to fermions
and bosons - precisely as possible and independent of any model or
assumptions - the
measurement of its total width is an important concern. The procedure
for ascertaining this quantity is based on the precisions for BF(H
\pfr ZZ$^*$) of 14.5 \% and the inclusive Higgsstrahlung cross
section of 2.8 \% . From that the error
of the total Higgs width is
calculated to $\Delta \Gamma_{tot}$(H) = $\pm$15 \%
which is mainly governed by the uncertainty of BF(H \pfr ZZ$^*$).

It is worthwile
to remind that for M$_H$ of about 120 GeV the
suggestion is \cite{gun} to measure the H \pfr \gaga 
partial width by means
of a \gaga collider \footnote {The H \pfr \gaga partial width itself
stands out as an observable of considerably physical importance. Recent
studies \cite{bossit} indicate that \gaga collisions allow 
for a $\sim$5 \% error determination of
$\Gamma$(H \pfr \gaga\hspace{-1mm}), for M$_H$ = 160 GeV.}, and if
combined with BF(H \pfr \gaga\hspace{-2mm}) and BF(H
\pfr \bb\hspace{-2mm}) measurements from \ee collisions,
$\Gamma_{tot}$(H) can be calculated.
An analogous procedure for a 160 GeV Higgs seems to be prevented as long
as an acceptable measurement fails for BF(H \pfr \gaga\hspace{-3mm}),
see sect.2.4. Fortunately, BF(H \pfr ZZ$^*$)
and $\sigma$(HZ) measurements allow to set a rather precise value for
$\Delta \Gamma_{tot}$(H) and, if in addition
the accurate value of BF(H \pfr WW$^*$) can be combined with a
precise measurement for the WWH coupling from e.g. the fusion reaction
\ee \pfr \nn H \pfr \nn WW$^{(*)}$, a superior total Higgs width is
feasible.

%----------------------------------------------------------------------
% sekt. 4
%----------------------------------------------------------------------
\section{Conclusions}

We have studied the prospects for measuring branching fractions
for a SM-like Higgs boson with M$_H$ = 160 GeV at the TESLA linear
\ee collider, assuming $\sqrt{s}$ = 350 GeV and an integrated
luminosity of 500 fb$^{-1}$. The resulting errors of these
measurements convoluted with an expected uncertainty of 2.8 \%
for the total Higgsstrahlung cross section are
summarized in Table \ref{tab:1}.

\begin{table}[h!t]
\begin{center}
\begin{tabular}{|l|l|}
\hline
& \\
Branching fraction & Expected error \\
\hline
& \\
BF (H \pfr WW$^*$) & \hspace{3mm}$\pm$ 3.5 \% \\
& \\
BF (H \pfr ZZ$^*$) & \hspace{3mm}$\pm$ 14.5 \% \\
& \\
BF (H \pfr \bb\hspace{-1mm}) & \hspace{3mm}$\pm$ 12 \% \\
& \\
\hline
\end{tabular}
\end{center}
\caption[ ] {\sl Branching fraction errors expected
  for the SM Higgs boson with M$_H$ = 160 GeV, at $\sqrt{s}$ = 350 GeV
  and an integrated luminosity of 500 fb$^{-1}$.}
\label{tab:1}
\end{table}
The measurement of BF(H \pfr \gaga\hspace{-2mm}) remains
a great challence since besides its very small SM
value itself the background expected to contribute is orders of magnitude
larger which renders a meaningful signal over background selection.

The procedure for ascertaining the total Higgs width and its error has been
outlined, with the result of $\Delta \Gamma_{tot}$(H) = $\pm$15 \%.
This estimate is based on uncertainties expected
for BF(H \pfr ZZ$^*$) of 14.5 \%
and the total Higgsstrahlung cross section of 2.8 \%.

Finally, we would like to emphasize that
the selection criteria proposed
are rather simple and not yet optimized. For the future,
we expect significant improvements if
the precise detector behaviour is known,
further topologies not yet involved in the analyses
are taken into account and, once the Higgs mass is known, $\sqrt{s}$
is optimized for running in the HZ measurement mode.

%----------------------------------------------------------------------
% Acknowledgments
%----------------------------------------------------------------------
\section*{Acknowledgments}

We would like to thank our colleagues within the course of physics
and detector studies of the
ECFA/DESY workshop series for many discussions and advices.
We also thank J.-C. Brient and R. Shanidze
for helpful discussions about the multidimensional
analysis.
E.B., V.I. and A.P. also thank the DESY-Zeuthen TESLA group for the
kind hospitality. Their work was partly supported by the
joint RFBR-DGF grant 99-02-04011.

%----------------------------------------------------------------------
% Bibliography
%----------------------------------------------------------------------

%\section*{References}

\end{document}